\documentclass[aps,prl,amsmath, twocolumn, groupedaddress]{revtex4}
\usepackage[T1]{fontenc}
\usepackage[latin9]{inputenc}
\usepackage{amstext}
\usepackage{graphicx}
\usepackage{amssymb}

\makeatletter

\providecommand{\LyX}{L\kern-.1667em\lower.25em\hbox{Y}\kern-.125emX\@}
\makeatother

\begin{document}

\preprint{}

\title{Calculation of Coupling Capacitance in Planar Electrodes}

\author{John M. Martinis$^1$, Rami Barends$^1$, and Alexander N. Korotkov$^2$}

\affiliation{$^1$Department of Physics, University of California, Santa Barbara, CA 93106, USA}

\affiliation{$^2$Department of Electrical Engineering, University of California, Riverside, CA 92521, USA}

\date{\today}

\begin{abstract}
We show how capacitance can be calculated simply and efficiently for electrodes cut in a 2-dimensional ground plane.  These results are in good agreement with exact formulas and numerical simulations.
\end{abstract}

\maketitle

The calculation of capacitance for complex electrode shapes is generally performed with numerical programs such as Sonnet or HFSS.  Long run times are typical because of the need to segment the electrodes into many elements, especially if ground planes are included.  Although capacitance is computed, there is usually little intuition gained as to how changing geometry will affect the result, except by running many iterations with different design parameters.

Here we show that for the geometry of a 2-dimensional ground plane with thin cuts that define the electrodes, the capacitance can be calculated with a simple formula.  For example, the coupling capacitance between two electrodes of arbitrary shape is given by
\begin{align}
C_{12} \simeq (\epsilon/\pi) A_1 A_2/r_{12}^3 \ ,
\end{align}
where $\epsilon$ is the average dielectric constant of the material above and below the ground plane, $A_1$ and $A_2$ are the areas of the electrodes 1 and 2, and $r_{12}$ is the distance between their centroids, assumed to be much greater than the extend of the electrodes.  Although this relation for the capacitance is formally exact only as the separation $s$ of the cuts go to zero, we show here that the scaling $s \rightarrow s/4$ allows accurate calculation of self capacitance using simple area integrals.

This work was motivated by the design of superconducting qubits, where it is necessary to set capacitance coupling elements between qubits, resonators, and control circuitry.  We believe this theory will also help with other design problems with integrated circuits, since computing capacitance in circuits with a ground plane is a common need.

In planar circuits, it is tempting to think that capacitance primarily arises at the electrode cuts.  In this note, it will be clear that this notion is incorrect, as a significant amount of capacitance comes from charge coupling to metal electrodes quite far from the cut.  The fundamental issue with planar circuits is that although there is a ground plane to shield the electric fields, the ground plane does not enclose all the electric fields coming from an electrode, so that the fields out of the plane give charge coupling at long distances.  This implies that stray coupling must be carefully considered and engineered so that these strays do not adversely affect device performance.

\section{Green's function solution}

We solve for capacitance using a Green function approach, which is based on the linearity of electromagnetism.  In general, the charge $dQ$ for a small area $dA_q$ is computed from a voltage source $V$ with area $dA_v$ using
\begin{align}
dQ/dA_q = f(r_q,r_v,\overrightarrow{r_c})\, V\,dA_v \ ,
\end{align}
where $f(r_q,r_v, \overrightarrow{r_c})$ is a function of the charge and voltage coordinates, as well as the positions $\overrightarrow{r_c}$ of all  other (infinitesimal) elements of the conductors.  The solution for $f$ is typically quite complex, as it can in general only be found using numerical techniques of matrix inversion, starting from a potential matrix having elements like $1/4\pi\epsilon|r_i-r_j|$.  However, for the simple geometry considered here of a ground plane sheet with infinitesimally thin cuts, this function turns out to be easy to calculate since the effects of the thin cuts can be neglected.

We thus need to find the Green's function solution for the charge distribution in an infinite ground plane coming from a infinitesimal electrode with a voltage $V$.  We first consider a simpler situation with a point source of charge $q$ at distance $d$ above the plane.  The problem can be solved using an image charge, which gives for radius $r$ and distance $z$ above the plane the potential
\begin{align}
U(r,z) &= \frac{q}{4\pi\epsilon} \Big[ \frac{1}{\sqrt{r^2+(z-d)^2}} - \frac{1}{\sqrt{r^2+(z+d)^2} } \Big] \\
&= \frac{M}{4\pi\epsilon} \frac{z}{(r^2+z^2)^{3/2} }  \ ,
\end{align}
where in the last equation we have made set $d \rightarrow 0$ but kept the total moment constant $M=2qd$.  The surface charge density in the ground plane is given by
\begin{align}
\sigma &= \epsilon \ \frac{\partial U(r,z)}{\partial z} \Big|_{z=0}
=-\frac{M}{4\pi r^3} \ . \label{eq:sigma}
\end{align}

We are interested in the charge distribution not from a charge dipole, but from an infinitesimal electrode of area $dA_v$ at voltage $V$.  The fields from this electrode can be equivalently described as arising from a spatial distribution of charge moments, which produce voltages equivalent to that coming from the electrode.  Although this distribution of moments could be found in principle, we need only solve a simpler problem of calculating the total moment, since we are only interested in the resulting surface charge density at large radius.

To proceed, we need to relate this total moment to the electrode voltage.  We first consider the simpler problem of a single dipole, where we sum the voltage coming from a dipole over the plane a distance $z$ above the ground
\begin{align}
\int U\,dA &= \int_0^\infty U(r,z)\,2\pi r dr \\
&=M/2\epsilon \, \label{eq:vmom}
\end{align}
where in the last equation we note the integral does not depend on the height $z$.  From linearity, we can relate the total moment $M_t$ to the integral of the voltage an infinitesimal distance above the ground plane, which gives
\begin{align}
M_t/2\epsilon = \int U\,dA
= V\, dV_v \label{eq:M-V}
\end{align}
since the ground plane at zero voltage gives no contribution to the integral.  Combining Eqs.\,(\ref{eq:sigma}) and (\ref{eq:M-V}) gives the charge distribution from the electrode at large radius
\begin{align}
\sigma = -\frac{\epsilon}{2\pi}\frac{ V dA_v}{r^3} \ .
\end{align}

The capacitance can now simply be calculated by summing the charge over all infinitesimal elements of the capacitors.   The general formula for capacitance between electrodes 1 and 2 is
\begin{align}
C_{12} &= \frac{\epsilon}{\pi}  \int \int \frac{dA_1\,dA_2}{|\mathbf{r}_1-\mathbf{r}_2|^3} \ , \label{eq:area}
\end{align}
where $\mathbf{r}_1$ and $\mathbf{r}_2$ are the coordinates of differential areas $dA_1$ and $dA_2$.

Here we have included capacitance from both above and below the ground plane, which increases the result by a factor of two.  For the case of a substrate and air with different dielectric constants $\epsilon_s$ and $\epsilon_a$, the capacitance is still given by the sum over the substrate and air capacitances, which is accounted for using the replacement $\epsilon \rightarrow (\epsilon_s + \epsilon_a)/2$.

\section{Comparison with coplanar lines}

\begin{figure}[t]
\includegraphics[scale=0.65]{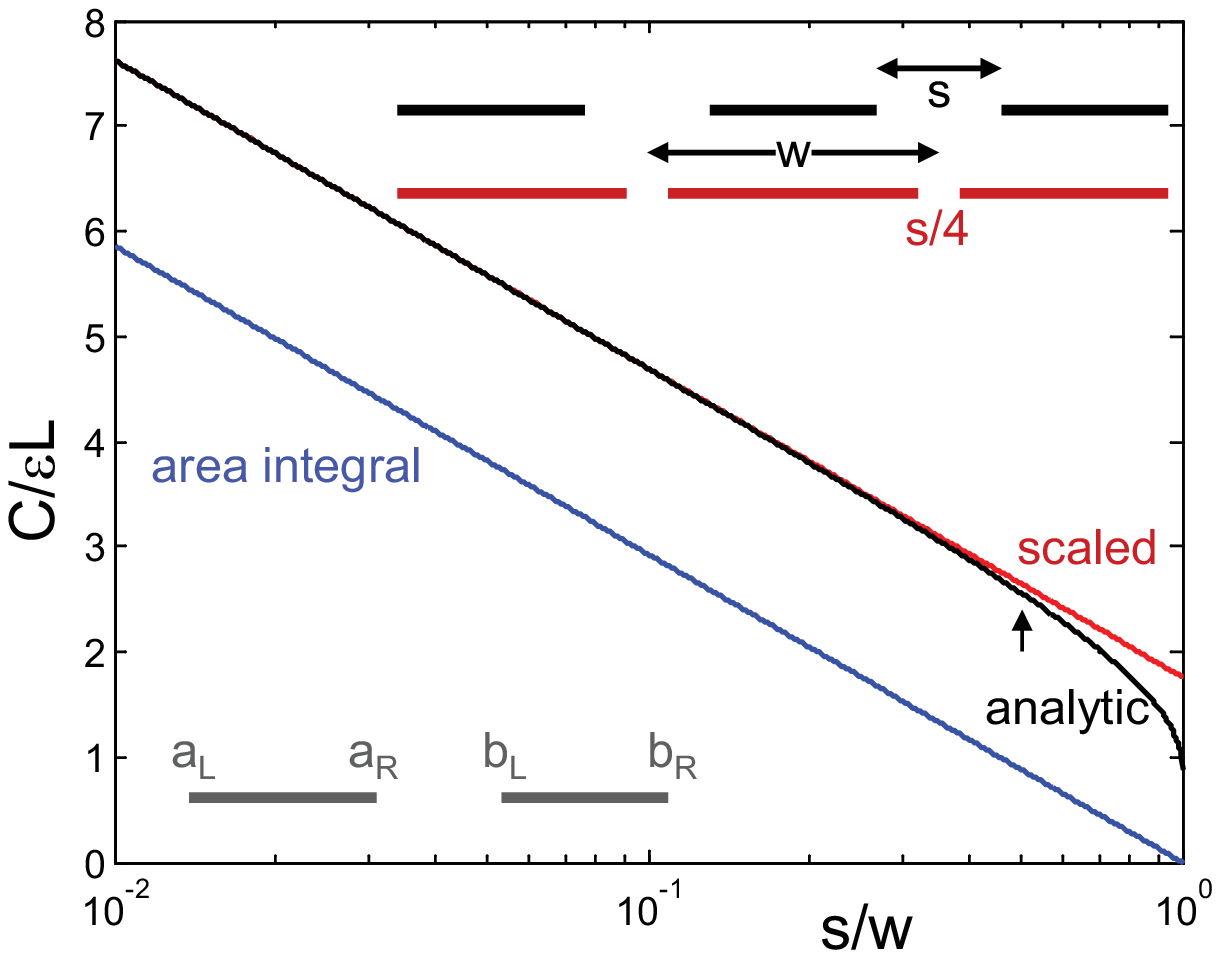}
\caption{\label{fig:capL} Plot of centerline capacitance versus $s/w$ for a coplanar waveguide of separation $s$, centered width $w$, and grounds extending to infinity. The analytical result Eq.\,(\ref{eq:CL}) is the black curve, whereas the blue line is the area integration formula Eq.\,(\ref{eq:CLint}).  Scaling of $s$ to $s/4$, illustrated in upper inset (red), shifts the integral prediction upward by $(4/\pi)\ln4$; when doing so, we find the analytical results are well matched even for small to moderate $s$.  Arrow indicates $s/w=0.5$.  The lower inset defines the coordinates $a_L, a_R, b_L$ and $b_R$ used in Eqs.\,(\ref{eq:cp2a}) and (\ref{eq:cp2b}).
}
\end{figure}

This theory may be tested against analytic results for the geometry of a coplanar waveguide with centered width $w$ and separation $s$, as illustrated in Fig.\,\ref{fig:capL}.  For this case the capacitance per unit length of the line is given by analytic results \cite{Gupta}
\begin{align}
C^{(cp,a)}/L &= 4\epsilon\,K(\kappa)/K(\sqrt{1-\kappa^2}) \\
& \simeq (4\epsilon/\pi) \ln[2(1+\sqrt{\kappa})/(1-\sqrt{\kappa})] \label{eq:CL} \\
&\simeq (4\epsilon/\pi) \ln(4w/s) \,
\label{eq:CLa}
\end{align}
where $\kappa=(w-s)/(w+s)$ and $K(\kappa)$ is the complete elliptic integral of the first kind.  The second equation is an excellent approximation for typical geometries, having errors greater than $10^{-2}$ only for $\kappa < 0.08$, whereas the last is valid for $s/w \rightarrow 0$.  This predicts a weak (logarithmic) dependence of capacitance on parameters of the line.

The capacitance from the area integral of Eq.\,(\ref{eq:area}) may be calculated directly for two coplanar electrodes $a$ ($b$), as illustrated in the lower inset of Fig.\,\ref{fig:capL}.  For left and right edge coordinates $a_L$ and $a_R$ ($b_L$ and $b_R$), the capacitance is
\begin{align}
C_{ab} &= (2\epsilon/\pi) \ln \Big[\frac{(a_L-b_L)(a_R-b_R)}{(a_R - b_L)(a_L-b_R)} \Big] \label{eq:cp2a} \\
&= (2\epsilon/\pi) \ln \Big[\,\frac{a_L-b_L}{a_R - b_L}\, \Big]
\ \ \ \ (\textrm{for } b_R \rightarrow \infty) \ . \label{eq:cp2b}
\end{align}

This expression can be readily be extended to the case of three electrodes.  For the coplanar line the self capacitance is
\begin{align}
C^{(cp)}/L = (4\epsilon /\pi) \ln(w/s) \label{eq:CLint}\ .
\end{align}

As shown in Fig.\,\ref{fig:capL}, the slope of this prediction matches the analytic formula Eq.\,(\ref{eq:CLa}), so the curves asymptotically match as $s \rightarrow 0$.  This justifies the idea that Eq.\,(\ref{eq:area}) is exact for separation $s \rightarrow 0$.  However, there is significant error for practical values of $s$ comparable in magnitude to $w$.  Figure\,\ref{fig:capL} also shows the analytic formula can be matched for $s \lesssim w/2$ by keeping $w$ constant but scaling $s \rightarrow s/4$: of course, this identification can be obtained by simply comparing Eqs.\,(\ref{eq:CLa}) and (\ref{eq:CLint}).  We conclude here that when computing self capacitance, Eq.\,(\ref{eq:area}) is only accurate for exponentially small separations, but a simple rescaling to $s/4$ corrects the formula quite well for edge effects.

\begin{figure}[t]
\includegraphics[scale=0.65]{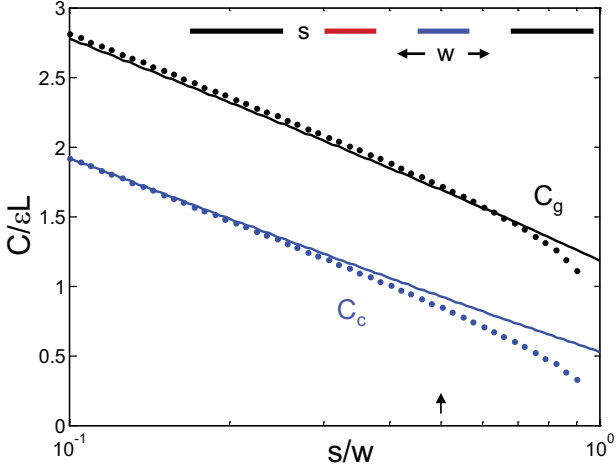}
\caption{\label{fig:CPS} Plot of ground capacitance $C_g$ (red to black electrodes) and coupling capacitance $C_c$ (red to blue) versus $s/w$.  Exact numerical results are points, whereas lines are results Eqs.\,(\ref{eq:2cpg}) and (\ref{eq:2cpc}) from area integration theory  with scaling $s\rightarrow s/4$.  Arrow indicates $s/w=0.5$.  The small shift between points and line are believed to arise from numerical errors.
}
\end{figure}
Although we can accurately calculate self capacitance for small separations, calculation of coupling capacitance should be even more reliable since the area elements are typically spaced by larger distance, so edge effects will be small.  To test this case, we next compare coplanar geometries with a second electrode.  For Fig.\,\ref{fig:CPS} where the two coplanar centerlines are separated by $s$, the ground and coupling capacitance is computed to be
\begin{align}
C_g^{(2cp)} &= \frac{2\epsilon}{\pi} \ln \Big( \frac{w}{s/4}\ \frac{2w}{w+s/4} \Big) \label{eq:2cpg} \\
C_c^{(2cp)} &= \frac{2\epsilon}{\pi} \ln\Big( \frac{w}{s/4}\,\frac{w}{2w-s/4} \Big) \ , \label{eq:2cpc}
\end{align}
where we have explicitly included the scaling $s \rightarrow s/4$.

\begin{figure}[t]
\includegraphics[scale=0.65]{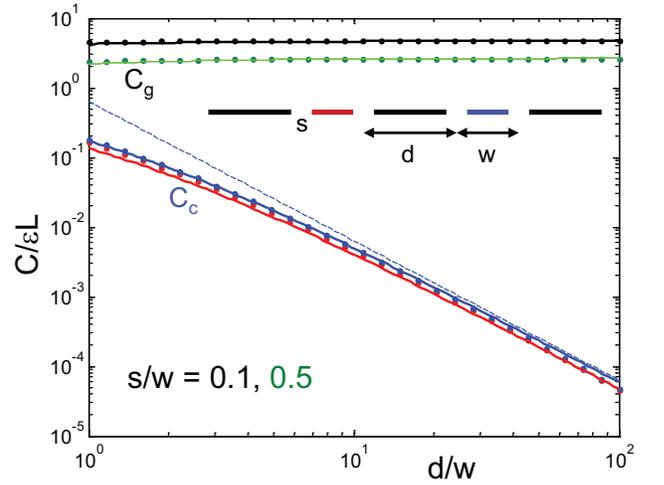}
\caption{\label{fig:CPD} Plot of capacitance versus $d/w$, as for Fig.\,\ref{fig:CPS} but with additional ground electrode of centered width $d$.  Theoretical predictions are from Eqs.\,(\ref{eq:dcpg}) and (\ref{eq:dcpc}), where black and blue (green and red) lines are for $s/w= 0.1\ (0.5)$.  The dashed line is the asymptotic prediction from Eq.\,(\ref{eq:2dpca}).
}
\end{figure}

For Fig.\,\ref{fig:CPD} where there is an additional intermediate ground electrode of length $d$, the capacitances are
\begin{align}
C_g^{(dcp)} &= \frac{2\epsilon}{\pi} \ln \Big(
\frac{w}{s/4}\ \frac{2w+d}{w+d+s/4}\ \frac{w}{s/4}\, \frac{d}{w+d-s/4}\Big)
\label{eq:dcpg} \\
C_c^{(dcp)} &=-\frac{2\epsilon}{\pi} \ln\Big[ 1-\Big( \frac{w-s/4}{d+w} \Big)^2 \Big] \label{eq:dcpc} \\
&\simeq \frac{2\epsilon}{\pi}  \Big( \frac{w}{d} \Big) ^2 \ \ \ \ \ (\textrm{for } s/4 \ll w \ll d ) \ .
\label{eq:2dpca}
\end{align}
Comparison of these formulas with an exact numerical solution shows excellent agreement.

\section{Useful Formulas}

\begin{figure}[t]
\includegraphics[scale=0.5]{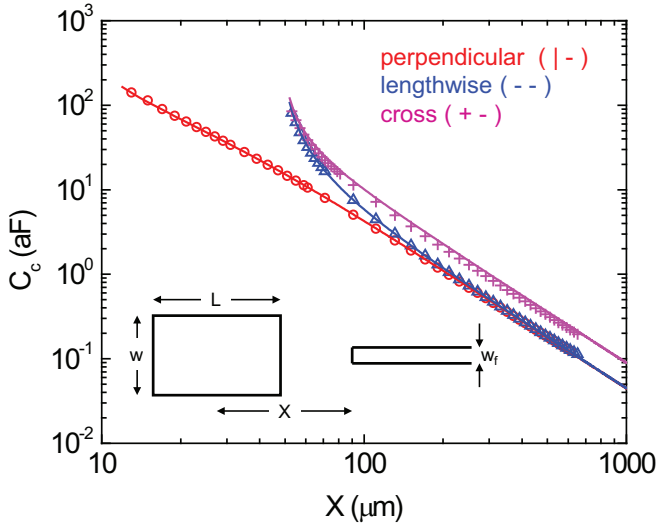}
\caption{\label{fig:numer1} Plot of coupling capacitance versus feedline distance $X$ for a rectangle with parallel or perpendicular orientations, and a cross.  Symbols are from Sonnet numerical simulations, whereas lines are obtained from the numerical integration formulas Eqs.\,(\ref{eq:rect}) or (\ref{eq:par}).  The rectangle has dimensions of length $\overline{L}=96\,\mu\textrm{m}$ and width $w=18\,\mu\textrm{m}$, and has a separation $s=2\,\mu\textrm{m}$ from electrode to ground.  The feedline has width $w_f=18\,\mu\textrm{m}$, and has its end a distance $X$ from the rectangle center.  Rectangle capacitance is $C=4.06\,\textrm{fF}$.
}
\end{figure}

We next show results for a feedline coupled to a rectangular electrode, as illustrated in Fig.\,\ref{fig:numer1}.  We consider a semi-infinite feedline of width $w_f$ coupled to a rectangular box of width $w=18\,\mu\textrm{m}$ and length $L=98\,\mu\textrm{m}$, separation $s$, and where the end of the feedline is separated from the center of the rectangle by distance $X$.  Here, we plot as points the results of  numerical calculations from Sonnet, for cases of parallel and perpendicular orientations.  We also show the predictions from the area integral formula Eq.\,(\ref{eq:area}), which gives
\begin{align}
C_c^{(r)} &= (\epsilon/\pi)\, w_f\,2
\Big[\ \frac{1}{y}\Big( \sqrt{x_+^2+y^2}-\sqrt{x_-^2+y^2}-L \Big)
\nonumber \\
&\ \ \ \ \ \ \ \ \ \ \ \ \ \ +\ln\Big( \frac{x_+}{x_-} \
\frac{\sqrt{x_-^2+y^2}+y}{\sqrt{x_+^2+y^2}+y} \Big) \ \Big] \ ,
\label{eq:rect}\\
x_\pm &= X \pm L/2 \ ,\\
y &= w/2 \ .
\end{align}
As shown in Fig.\,\ref{fig:numer1}, we see excellent agreement between this formula and the numerical data.  For the case $w \ll L$ the geometry is for two colinear lines, and we use a line integral version of Eq.\,(\ref{eq:area}) to find
\begin{align}
C_c^\|   &= (\epsilon/\pi)\, w_f w\, \frac{L/2}{X^2-(L/2)^2}
 \ \ \ \ (\textrm{for } \omega \ll X )
\label{eq:par} \\
&\simeq (\epsilon/\pi)\, \,\frac{(w_f X/2) (wL)}{X^3}\ \ \ \ \ (\textrm{for } \omega, L \ll X ) \ \label{eq:rectsm}.
\end{align}
For the perpendicular case, we find.
\begin{align}
C_c^\bot   &= (\epsilon/\pi)\, \frac{4w_f w}{L}\, [\sqrt{1+(L/2X)^2}-1]
 \ \ \ \ (\textrm{for } \omega \ll X )
\label{eq:perp} \ .
\end{align}

Note that Eq.\,(\ref{eq:rectsm}) is written as the product of areas divided by a distance cubed to match the form of the area integral formula.  Here, it is seen that the effective length of the feedline is half the spacing $X$.  We will express the following formulas in this manner to emphasize this basic form.

For a rectangle separated from a small area $A$ with center to center distance $X$, the coupling capacitance is
\begin{align}
C_c^{(rA)}   &= \frac{\epsilon}{\pi}\frac{2A}{y} \Big[ \sqrt{1+y^2/x_-^2}-\sqrt{1+y^2/x_+^2}\, \Big] \\
&=\frac{\epsilon}{\pi}\frac{A(wL)}{X^3} \Big[ 1-\frac{1}{8}\Big(\frac{w}{X}\Big)^2+\frac{1}{2}\Big(\frac{L}{X}\Big)^2 + ...\ \Big]
\end{align}
As the correction terms are second order in $w$ and $L$, there is less than 10\% change in the simple area formula even for relatively large size $w \le 0.9\,X$ and $L \le 0.45 X$.

For the condition $L \rightarrow \infty$, Eq.\,(\ref{eq:par}) can be used to calculate the coupling capacitance between two colinear semi-infinite lines separated by a gap $g$, giving
\begin{align}
C_c^{\| g}   &= (\epsilon/\pi)\frac{(w_f g)(w g)}{2g^3} \ .
\end{align}
For the case of perpendicular geometry with coupling between a semi-infinite and infinite line separated by gap $g$, we find
\begin{align}
C_c^{\bot g}   &= (\epsilon/\pi)\frac{2(w_f g )(w g)}{g^3}\ .
\end{align}
For the case of an infinite line coupled to a rectangle of width $w$ and length $L$, parallel to the infinite line but offset by a gap $g$, the coupling capacitance is
\begin{align}
C_c^\textrm{offset}   &= (\epsilon/\pi)\frac{2 w_f(w L)}{g^2} \ .
\end{align}

\section{Comparison of square and rectangle}

Isolation of capacitance is an important issue, so an obvious question is whether the electrode geometry of a square or rectangle gives better isolation from one qubit to another.  Since the coupling capacitance scales as the area of each electrode $A$, but the self capacitance scales as their perimeter $p$, the ratio of the coupling to self capacitance is $A/p$.  Assuming equal self capacitance, this result implies that the lowest stray coupling will come from a rectangle of small aspect ratio $w/L$.  Smaller coupling arises because the ground plane around the narrow dimension of the rectangle better screens the electric fields.

\section{Finite substrate thickness}

The integral solution developed in this paper can be also applied to other important cases, such as for a metal plane on top of a substrate of finite thickness $t$ and dielectric constant $\epsilon_s$.  Since the symmetry in the x and y directions is preserved, we expect a Green's function with only a radial dependence $f(r)/r^3$, which has $f(r) = 1$ for $r \ll t$, but then changes for $r \gg t$ due to screening effects.

\begin{figure}[t]
\includegraphics[scale=0.68]{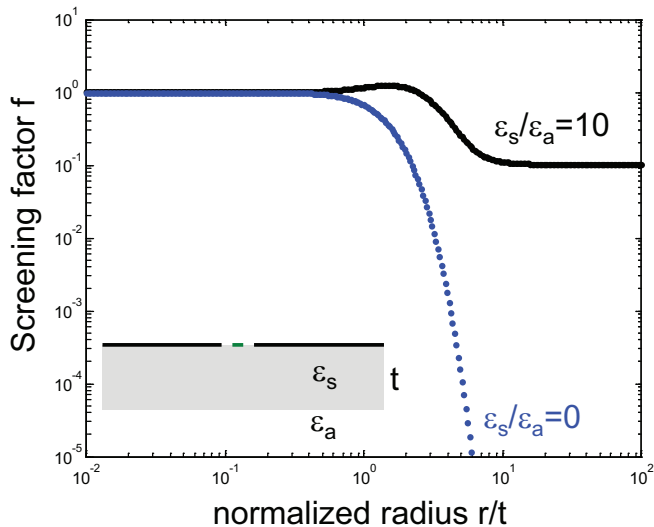}
\caption{\label{fig:sb} Plot of screening factor $f$ versus normalized radius $r/t$ for a metal plane on top of a substrate of thickness $t$ and dielectric constant $\epsilon_s$.  Underneath the substrate is another dielectric with $\epsilon_a$.  We consider the substrate-air case with $\epsilon_s/\epsilon_a=10$ (black), as well as for a metal ground plane underneath the substrate $\epsilon_a \rightarrow \infty$ (blue).  }
\end{figure}

We consider the geometry shown in the inset of Fig.\,\ref{fig:sb}, where a dielectric with $\epsilon_a$ is below the substrate.  For the case of an air dielectric, we expect $\epsilon_a=1$, whereas the case of the substrate on a metal ground plane can be solved using $\epsilon_a \rightarrow \infty$.

The radial scaling of the Green function can be solved by considering a charge source $q$ just below the origin.  The metal plane gives an image charge of $-q$ just above the origin, whereas the $\epsilon_s$-$\epsilon_a$ boundary produces an image charge at the z-coordinate $-2t$ with charge $q'\equiv -\alpha\,q$ and $\alpha = (\epsilon_a-\epsilon_s)/(\epsilon_a+\epsilon_s)$.  The effect of the two boundaries is to produce a series of image dipoles in the z direction each spaced by $2t$, with moments that are repeatedly reduced by the factor $\alpha$ away from the origin.  Since the charge on the metal plane is proportional to the electric field in the z-direction from these dipoles, the screening factor is given by a sum $n$ over the images
\begin{align}
f(r) = 1+2 \sum_{n=1,2}^{\infty}
\alpha^n \frac{r^3}{[r^2+(2tn)^2]^{3/2}}
\Big[ 1-\frac{3}{1+(r/2tn)^2} \Big]
\end{align}

This screening factor $f(r)$ is plotted in Fig.\,\ref{fig:sb} for the case of a substrate-air interface with $\epsilon_s/\epsilon_a = 10$ ($\alpha=-9/11$), and for a metal ground plane below the substrate $\epsilon_s/\epsilon_a = 0$ ($\alpha = 1$).  For the air case, the screening factor reduces to 1/10 for $r \gg t$, as expected since the effective dielectric constant should be that of air.  For the metal case, the screening factor drops rapidly to zero, as expected for charge screening.

The summation generally converges taking a maximum $n$ of about $10^4$. However, the sum is slowly convergent for the case $\alpha = 1$ and $r/t > 5$, so one should then use the fit function $f(r/t) = f(5)\,\exp[-2.718(r/t-5)]$.  For practical implementation, it is suggested to compute $f(r)$ once and then use an interpolation function for the area integration.

The total charge on the metal plane comes from electric fields above and below the substrate.  This can be directly summed, which changes the integral of Eq.\,(\ref{eq:area}) by the following replacement
\begin{align}
\frac{\epsilon}{r^3} \rightarrow \frac{\epsilon_a}{2r^3} + \frac{\epsilon_s}{2r^3}f(r) \ . \label{eq:subs}
\end{align}

An interesting calculation considers the coupling capacitance to the substrate ground plane, which we take as the charge $[1-f(r)]/r^3$ that is no longer accumulating at the top plane.  For a source disk of radius $R \ll t$ and area $A$, we find
\begin{align}
C_{gz} &= \frac{\epsilon_s  A}{2\pi} \int_R^\infty \frac{2\pi r }{r^3}[1-f(r)] \,dr\\
&=\frac{\epsilon_s  A}{t} \label{eq:cgz}
\end{align}
where we have ignored the rise in charge density at $r \simeq R$ since $1-f(r) \simeq 0$ for $r \ll t$.  It is possible to simply estimate this result by computing an integral with a step-function cutoff to the coupling for $r > t$, which also gives the same result. This equation shows a relatively slow dependence  $1/t$ on the thickness, so capacitance to this ground can not obviously be ignored.

Equation\,(\ref{eq:cgz}) was derived for a disk of small area $A$, so it represents the Green function for this ground-plane geometry.  For an electrode of any area and shape, the ground capacitance $C_{gz}$ can be obtained by integrating over all the area, also giving the result of Eq.\,(\ref{eq:cgz}) where now $A$ is the area of an arbitrary electrode.  As expected, the ground capacitance is the parallel plate result $\epsilon_s A/t$ for large area, where the extent of the electrode is much larger than $t$.  Perhaps surprisingly, the parallel plate formula is also good for the case of small area.

This parallel plate result can be used to estimate capacitance from more complicated geometries.  For example, if the dielectric stackup was substrate, air and then a ground plane, the formula for the ground capacitance would simply be the parallel plate formula corresponding to the substrate and air.

\end{document}